# Modified extended tanh-function method and nonlinear dynamics of microtubules

Slobodan Zdravković[a,*], Louis Kavitha[b,c], Miljko V. Satarić[d], Slobodan Zeković[a], Jovana Petrović[a]

[a] *Institut za nuklearne nauke Vinča, Univerzitet u Beogradu, Poštanski fah 522, 11001 Beograd, Serbia*
[b] *Department of Physics, Periyar University, Salem-636 011, India*
[c] *The Abdus Salam International Centre for Theoretical Physics, Trieste, Italy*
[d] *Fakultet tehničkih nauka, Univerzitet u Novom Sadu, 21000 Novi Sad, Serbia*

A B S T R A C T

We here present a model of nonlinear dynamics of microtubules (MT) in the context of modified extended tanh-function (METHF) method. We rely on the ferroelectric model of MTs published earlier by Satarić et al [1] where the motion of MT subunits is reduced to a single longitudinal degree of freedom per dimer. It is shown that such nonlinear model can lead to existence of kink solitons moving along the MTs. An analytical solution of the basic equation, describing MT dynamics, was compared with the numerical one and a perfect agreement was demonstrated. It is now clearer how the values of the basic parameters of the model, proportional to viscosity and internal electric field, impact MT dynamics. Finally, we offer a possible scenario of how living cells utilize these kinks as signaling tools for regulation of cellular traffic as well as MT depolymerisation.

## 1. Introduction

Microtubules are major cytoskeletal proteins. They are hollow cylinders formed by protofilaments (PF) representing series of proteins known as tubulin dimers [2,3]. There are usually 13 longitudinal PFs covering the cylindrical walls of MTs. The inner and the outer diameters of the cylinder are 15nm and 25nm, while its length may span dimensions from the order of micrometer to the order of millimetre. Each dimer is an electric dipole whose length and longitudinal component of the electric dipole moment are $l = 8$nm [2-4] and $p = 337$Debye [5], respectively. The constituent parts of the dimers are $\alpha$ and $\beta$ tubulins, corresponding to positively and negatively charged sides, respectively [2-4].

[*]Corresponding author.
 E-mail addresses: szdjidji@vinca.rs (S.Zdr.), louiskavitha@yahoo.co.in (L.K.), bomisat@neobee.net (M.V.S.), zekovic@vinca.rs (S.Zek.), jovanap@vinca.rs

In this paper we demonstrate how METHF method [6-10] can be used in the study of nonlinear dynamics of MTs. The paper is organized as follows. In Section 2 we explain the well known model for MTs we rely on [1]. The modification of the model presented in this paper is a generalization of the original one and will be referred to as u-model. The model brings about a crucial nonlinear differential equation, describing nonlinear dynamics of MTs. In Section 3 we briefly describe METHF method. Then we solve the basic nonlinear differential equation, mentioned above. We show that its solution is a kink-like solitonic wave. This result is compared with numerical solutions in Section 4. Finally, in Section 5, we give concluding remarks. In particular, we emphasize the biological importance of the studied kink-like solitons.

## 2. U-model of MTs

The model we rely on assumes only one degree of freedom of dimers motion within the PF [1]. This is a longitudinal displacement of a dimer at a position n denoted as $u_n$ and thus we call the model as u-model.

The overall effect of the surrounding dimers on a dipole at a chosen site n can be described by a double-well potential [1]

$$V_d(u_n) = -\frac{1}{2}Au_n^2 + \frac{1}{4}Bu_n^4 \qquad (1)$$

where $A$ and $B$ are positive parameters that should be estimated. As an electrical dipole, a dimer in the intrinsic electric field of the MT acquires the additional potential energy given by [1]

$$V_{el}(u_n) = -Cu_n, \qquad C = qE, \qquad (2)$$

where $E$ is the magnitude of the intrinsic electric field at the site n, as the dimer n exists in the electric field of all other dimers, and $q$ represents the excess charge within the dipole. It is assumed that $q > 0$ and $E > 0$.

The Hamiltonian for one PF is represented as follows

$$H = \sum_n \left[ \frac{m}{2}\dot{u}_n^2 + \frac{k}{2}(u_{n+1} - u_n)^2 + V(u_n) \right], \qquad (3)$$

where dot means the first derivative with respect to time, $m$ is a mass of the dimer, $k$ is an intra-dimer stiffness parameter and the integer $n$ determines the position of the considered dimer in the PF [1]. The first term represents a kinetic energy of the dimer, the second one is a potential energy of the chemical interaction between the neighbouring dimers belonging to the same PF and the last term is the combined potential

$$V(u_n) = -Cu_n - \frac{1}{2}Au_n^2 + \frac{1}{4}Bu_n^4. \qquad (4)$$

It is obvious that the nearest neighbour approximation is used. However, this does not mean that the influence of the neighbouring PFs is completely ignored as the value of the electric field $E$ depends also on the dipoles belonging to the neighbouring PFs.

By using the generalized coordinates $q_n$ and $p_n$, defined as $q_n = u_n$ and $p_n = m\dot{u}_n$, applying a continuum approximation $u_n(t) \to u(x,t)$ and making a series expansion

$$u_{n\pm 1} \to u \pm \frac{\partial u}{\partial x}l + \frac{1}{2}\frac{\partial^2 u}{\partial x^2}l^2 \qquad (5)$$

we can straightforwardly obtain an appropriate dynamical equation of motion. In order to derive a realistic equation, the viscosity of the solvent should also be taken into consideration. This can be achieved by introducing a viscosity force $F_v = -\gamma \dot{u}$ into the obtained dynamical equation of motion, where $\gamma$ is a viscosity coefficient [1]. All this brings about the following nonlinear partial differential equation

$$m\frac{\partial^2 u}{\partial t^2} - kl^2\frac{\partial^2 u}{\partial x^2} - qE - Au + Bu^3 + \gamma\frac{\partial u}{\partial t} = 0. \qquad (6)$$

It is well known that, for a given wave equation, a travelling wave $u(\xi)$ is a solution which depends upon $x$ and $t$ only through a unified variable $\xi$

$$\xi \equiv \kappa x - \omega t, \qquad (7)$$

where $\kappa$ and $\omega$ are constants. Substitution of $x$ and $t$ by $\xi$ transforms Eq. (6) into the following ordinary differential equation (ODE)

$$(m\omega^2 - kl^2\kappa^2)u'' - \gamma\omega u' - Au + Bu^3 - qE = 0. \qquad (8)$$

By introducing a dimensionless function $\psi$ through the relation

$$u = \sqrt{\frac{A}{B}}\psi, \qquad (9)$$

a much more convenient equation can be obtained. This final ODE reads

$$\alpha\psi'' - \rho\psi' - \psi + \psi^3 - \sigma = 0, \qquad (10)$$

and contains the following new parameters:

$$\alpha = \frac{m\omega^2 - kl^2\kappa^2}{A}, \qquad (11)$$

$$\sigma = \frac{qE}{A\sqrt{\dfrac{A}{B}}}, \qquad (12)$$

$$\rho = \frac{\gamma\omega}{A} \qquad (13)$$

and $u' \equiv \dfrac{du}{d\xi}$.

It was already mentioned that this approach represented a certain improvement of the original model, explained in [1]. If we compare Eq. (10) with the appropriate one in [1] we can see that they are equal for $\alpha = -1$. Therefore, this approach is more general. We treat the parameters $\rho$ and $\sigma$ as an input and will determine values of dinamical parameters of the system. We will see that the final result depends on $\rho$ and $\sigma$ only, i.e. on the parameters that determine their values.

The crucial equation (10) will be solved in the next section. Before we proceed we want to discuss the potential energy $V(u)$, defined by Eq. (4). This step is very important to understand the physics behind Eq. (10) and its solutions. Using the procedure explained above we can easily obtain the following convenient expression for this potential

$$V(\psi) = \frac{A^2}{B} f(\psi), \qquad (14)$$

where

$$f(\psi) = -\sigma\psi - \frac{1}{2}\psi^2 + \frac{1}{4}\psi^4. \qquad (15)$$

The function $f(\psi)$ is shown in Fig. 1 for two values of the parameter $\sigma$. For $\sigma = 0$ the function $f(\psi)$ and, consequently, the potential $V(\psi)$, is symmetric (curve a) while for the increasing $\sigma$ the right minimum becomes deeper and the left one becomes shallower and elevated. To find the values of $\psi$ for which $f(\psi)$ reaches a maximum and minima we should solve the equation

$$f'(\psi) = -\sigma - \psi + \psi^3 = 0. \qquad (16)$$

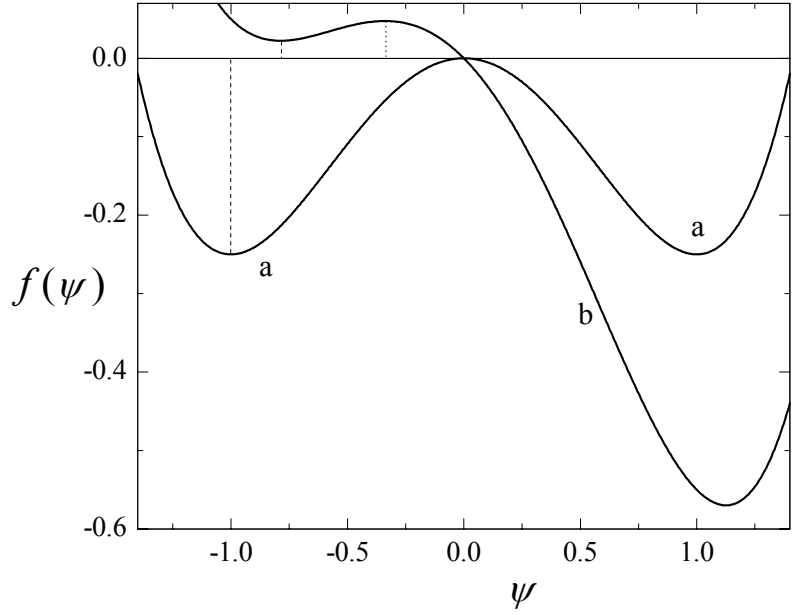

**Fig. 1.** The function $f(\psi)$ for: (a) $\sigma = 0$ and (b) $\sigma = 0.3$.

According to the procedure explained in Appendix we can easily obtain the following roots of Eq. (16):

$$\psi_R = \frac{2}{\sqrt{3}} \cos F, \tag{17}$$

$$\psi_{max} = \frac{1}{\sqrt{3}} \left( -\cos F + \sqrt{3} \sin F \right), \tag{18}$$

$$\psi_L = -\frac{1}{\sqrt{3}} \left( \cos F + \sqrt{3} \sin F \right), \tag{19}$$

where

$$F = \frac{1}{3} \arccos \left( \frac{\sigma}{\sigma_0} \right). \tag{20}$$

The functions $\psi_R$ and $\psi_L$ correspond to the right and left minimum of the function $f(\psi)$ and a critical value of the parameter $\sigma$ has the value

$$\sigma_0 = \frac{2}{3\sqrt{3}}. \tag{21}$$

This means that the three real roots of Eq. (16) exist for $\sigma < \sigma_0$.

The dependence $\psi_{ex}(\sigma)$ is shown in Fig. 2, where $\psi_{ex}$ stands for $\psi_R$, $\psi_{max}$ and $\psi_L$. We see that $\psi_R$ is a slowly increasing function of $\sigma$ while $\psi_L$ and $\psi_{max}$ approach each other for the increasing $\sigma$. This means that the right minimum moves to the right for the higher $\sigma$ while the maximum and the left minimum of the function $f(\psi)$ become closer. All this can be noticed in Fig. 1. For $\sigma = \sigma_0$ the left minimum and the maximum disappear, coalescing in the saddle point.

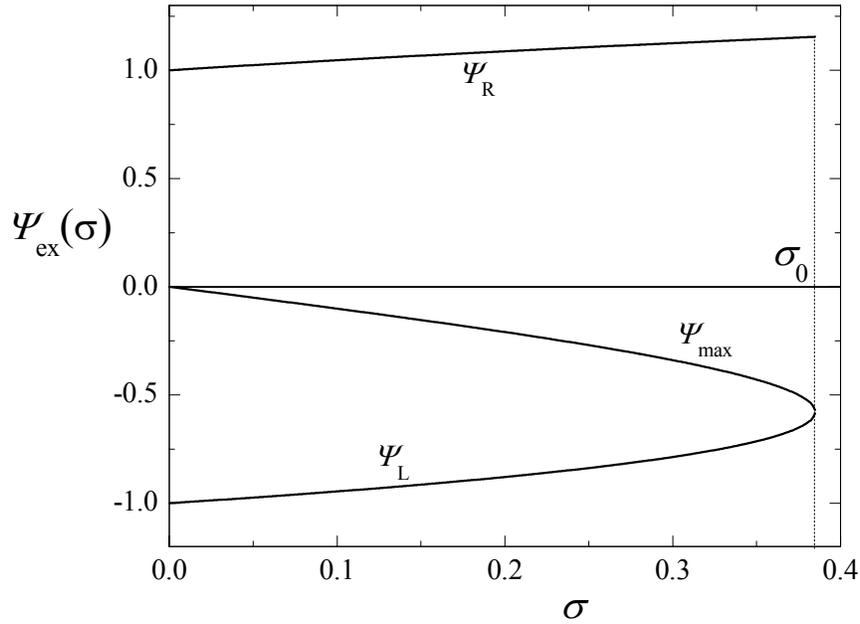

**Fig. 2.** The values of $\psi_{ex}$ corresponding to the extrema of the potential $V(\psi)$ as a function of the parameter $\sigma$.

The values $\psi_R$ and $\psi_{max}$ for a couple of values of $\sigma$ are shown in Table 1. Of course, $\psi_{max}$ does not exist for $\sigma = \sigma_0$.

**Table 1**

| $\sigma$ | 0 | 0.1 | 0.2 | 0.3 | $\sigma_0$ |
|---|---|---|---|---|---|
| $\psi_R$ | 1 | 1.047 | 1.088 | 1.125 | 1.155 |
| $\psi_{max}$ | 0 | -0.101 | -0.209 | -0.339 | n a |

All this suggests that the higher $\sigma$, i.e. the bigger value of $qE$, increases stability of MTs dynamics around the right minimum which becomes deeper. This requires further research and should be checked by stability analysis which will be a topic of a separate publication.

## 3. Modified extended tanh-function method

In what follows we briefly outline the basic features of METHF method. The method will be applied for solving Eq. (10). According to this procedure we look for the possible solution in the form [6,7]

$$\psi = a_0 + \sum_{i=1}^{M}\left(a_i \Phi^i + b_i \Phi^{-i}\right), \tag{22}$$

where the function $\Phi = \Phi(\xi)$ is a solution of the well known Riccati equation

$$\Phi' = b + \Phi^2 \tag{23}$$

and $\Phi'$ is the first derivative [6,7]. The parameters $a_0$, $a_i$, $b_i$ and $b$ are real constants that should be determined as well as an integer $M$. The possible solutions of (23) depend on the parameter $b$ as follows:

1) If $b > 0$ then $\Phi = \sqrt{b}\tan(\sqrt{b}\xi)$, or $\Phi = -\sqrt{b}\cot(\sqrt{b}\xi)$, (24)
2) If $b = 0$ then $\Phi = -\dfrac{1}{\xi}$, (25)
3) If $b < 0$ then $\Phi = -\sqrt{-b}\tanh(\sqrt{-b}\xi)$, or $\Phi = -\sqrt{-b}\coth(\sqrt{-b}\xi)$. (26)

Balancing the orders of $\psi''$ and $\psi^3$ with respect to the new function $\Phi$ we find $M = 1$. Namely, the highest orders of the function $\Phi$ in the expressions for $\psi''$ and $\psi^3$ are $\Phi^{M+2}$ and $\Phi^{3M}$ respectively and they are equal for $M = 1$.

Eqs. (22) and (23) yield the expressions for $\psi'$, $\psi''$ and $\psi^3$ as functions of $\Phi$. For example, the second derivative of $\psi$ is

$$\psi'' = 2a_1b\Phi + 2a_1\Phi^3 + 2b_1b^2\Phi^{-3} + 2b_1b\Phi^{-1}. \tag{27}$$

Eq. (10) in terms of the expressions for $\psi$, $\psi'$, $\psi''$ and $\psi^3$ brings about the crucial equation

$$A_1\Phi + B_1\Phi^{-1} + A_2\Phi^2 + B_2\Phi^{-2} + A_3\Phi^3 + B_3\Phi^{-3} + A_0 = 0, \tag{28}$$

where the following set of abbreviations is introduced:

$$A_0 = -a_0 + a_0^3 + 6a_0a_1b_1 - \rho a_1b + \rho b_1 - \sigma, \tag{29}$$

$$A_1 = -a_1 + 3a_0^2a_1 + 2\alpha a_1b + 3a_1^2b_1, \tag{30}$$

$$B_1 = -b_1 + 3a_0^2b_1 + 2\alpha bb_1 + 3a_1b_1^2, \tag{31}$$

$$A_2 = 3a_0a_1^2 - \rho a_1, \tag{32}$$

$$B_2 = 3a_0b_1^2 + \rho bb_1, \tag{33}$$

$$A_3 = 2\alpha a_1 + a_1^3 \tag{34}$$

and

$$B_3 = 2\alpha b^2b_1 + b_1^3. \tag{35}$$

Of course, Eq. (28) is satisfied if all these coefficients are simultaneously equal to zero which renders a system of seven equations. Before embarking on solving such a system, we examine its behaviour under the conditions given by Eqs. (24)-(26). The solutions expressed through tangents and cotangents cannot be biophysically tractable as these functions diverge. The same can be said for hyperbolic cotangent. This means that we are looking for the acceptable solutions for which $b < 0$, $a_1 \neq 0$ and $b_1 = 0$, which reduces the system mentioned above. Hence, the system can be reduced to the following system of four equations

$$\left.\begin{array}{l} -a_0 + a_0^3 - \rho a_1b - \sigma = 0, \\ -1 + 3a_0^2 + 2\alpha b = 0, \\ \rho = 3a_0a_1, \\ 2\alpha = -a_1^2. \end{array}\right\} \tag{36}$$

Its solutions, i.e. the values of the parameters $b$, $a_0$, $a_1$ and $\alpha$, are given through

$$8a_0^3 - 2a_0 + \sigma = 0, \tag{37}$$

$$a_1 = \frac{\rho}{3a_0}, \tag{38}$$

$$\alpha = -\frac{a_1^2}{2} \tag{39}$$

and

$$b = \frac{3a_0^2 - 1}{a_1^2}. \tag{40}$$

Notice that $a_0 a_1 > 0$ and $\alpha < 0$. Also, $b < 0$ holds for the inequality

$$a_0^2 < \frac{1}{3}. \tag{41}$$

It is easy to check that the requirement (41) is equivalent to $\sigma < \sigma_0$, which was discussed earlier.

It was mentioned above that this extended version of the model reduces to the model from [1] if $\alpha = -1$. To calculate the value for $\alpha$ we have to know the values of the parameters determining $\sigma$ and $\rho$. The values of these parameters have not been determined yet. It suffices now to discuss a negative sign of $\alpha$, as can be seen from (39). This parameter comes from the first two terms in Eq. (6). These are inertial and elastic terms. A question is which contribution is higher and the negative value of $\alpha$ suggests that the elastic term in (6) is bigger than the inertial one. Using the expression for wave speed

$$v = \frac{\omega}{\kappa} \tag{42}$$

Eq. (11) can be written as

$$\alpha = \frac{\kappa^2}{A}\left(mv^2 - kl^2\right). \tag{43}$$

The fact that $\alpha < 0$ may indicate a small velocity of the wave and/or a big $k$, i.e. strong chemical bond between the neighbouring dimers in PF, as can be seen from (3).

The importance of $\alpha$ and our intention to get as much information about it as possible is reason that we did not perform rescaling of Eq. (10), which would remove $\alpha$ from the equation. Notice that we cannot *a priori* exclude the value $\alpha = 0$, which prevents rescaling.

Eq. (37) looks like (16) and Appendix should be used again. The requirement for the existence of the three real solutions is $\sigma < \sigma_0$ again. Hence, the roots of (37) are

$$a_{01} = \frac{1}{2\sqrt{3}}\left(\cos F + \sqrt{3}\sin F\right), \tag{44}$$

$$a_{02} = \frac{1}{2\sqrt{3}}\left(\cos F - \sqrt{3}\sin F\right), \tag{45}$$

$$a_{03} = -\frac{1}{\sqrt{3}}\cos F \tag{46}$$

where the expressions for $F$ and $\sigma_0$ are given by (20) and (21). Of course, all these three values depend on $\sigma$ and they are shown in Fig. 3. One can see that (37) has three real solutions for $\sigma < \sigma_0$ and only one ($a_{03}$) for $\sigma > \sigma_0$. To obtain Eqs. (44)-(46) a relationship

$$\cos^{-1}(\beta) + \cos^{-1}(-\beta) = \pi \tag{47}$$

was used.

A next step is to construct the function $\psi(\xi)$. Suppose that $\psi_1$ describes the system in the state corresponding to the right minimum in Fig. 1. Using Eqs. (26), (38) and (40) we easily obtain

$$\Phi = -\sqrt{-b}\,\tanh\left(\sqrt{-b}\,\xi\right) = -\frac{3a_{01}}{\rho}\sqrt{1-3a_{01}^2}\,\tanh\left(\frac{3a_{01}}{\rho}\sqrt{1-3a_{01}^2}\,\xi\right). \tag{48}$$

Finally, Eqs. (22) with $M = 1$ and (48) yield the following expression for the function $\psi_1$:

$$\psi_1(\xi) = a_{01} - \sqrt{1-3a_{01}^2}\,\tanh\left(\frac{3a_{01}}{\rho}\sqrt{1-3a_{01}^2}\,\xi\right), \tag{49}$$

where $a_{01}$ is determined by (44), (20) and (21). We can see that $\psi_1(\xi)$ depends on $\rho$ and $\sigma$ only as $a_{01}$ is a function of $\sigma$. Notice that the jump of the function $\psi_1(\xi)$ from $-\infty$ to $+\infty$ depends on $\sigma$ only while the solitonic width, i.e. slope, depends on both $\rho$ and $\sigma$. It is obvious that the solitonic width is proportional to viscosity.

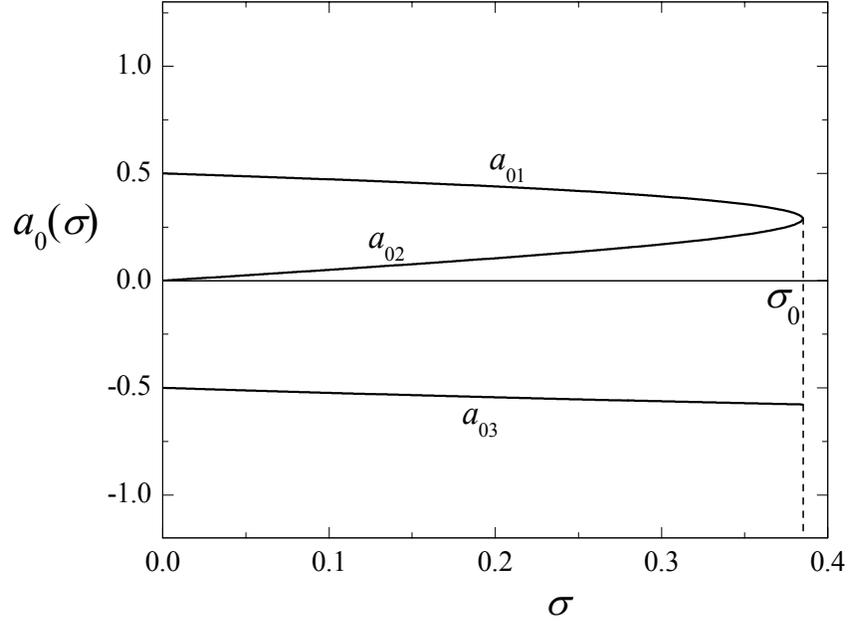

**Fig. 3.** The values of $a_0$ as a function of the parameter $\sigma$.

The function (49) is shown in Fig. 4 (analytic curves) for the two values of the parameter $\sigma$ and for $\rho = 1$. Obviously, this is an anti-kink soliton. The parameter $\rho$ affects its slope, i.e. the width of the soliton, but not the character of the wave.

According to Eq. (49) we can calculate the initial and the final value of $\psi_1$ that is

$$\psi_1(-\infty) = a_{01} + \sqrt{1 - 3a_{01}^2} \tag{50}$$

and

$$\psi_1(+\infty) = a_{01} - \sqrt{1 - 3a_{01}^2} \; . \tag{51}$$

Using Eq. (44) we can straightforwardly prove that

$$\psi_1(-\infty) = \psi_R, \qquad \psi_1(+\infty) = \psi_{max} \tag{52}$$

as $\sin F - \sqrt{3} \cos F$ is negative.

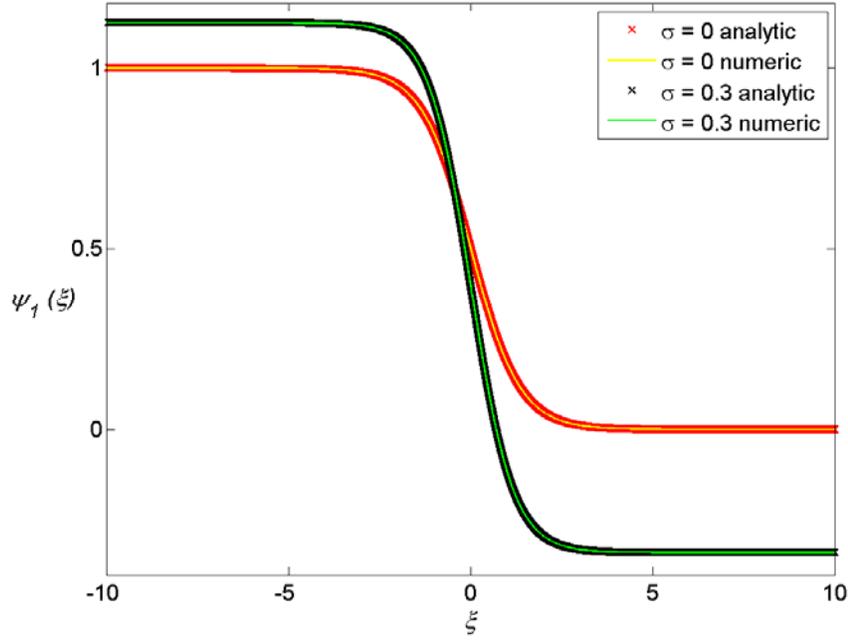

**Fig. 4.** The function $\psi_1(\xi)$ for $\rho = 1$ and: (a) $\sigma = 0$ and (b) $\sigma = 0.3$.

Therefore, nonlinear dynamics of MTs is described by antikink solitons. In other words, the system exists at the right minimum of Fig. 1 and the value of $\psi(\xi)$ is $\psi_1(-\infty) = \psi_R$. When some energy is supplied, released by hydrolysis of guanosine triphosphate (GTP), the system jumps to the maximum and $\psi$ becomes $\psi_{max}$, as explained by (52). The way how this transition is performed is explained by the solitonic wave (49). Of course, sooner or later, the system spontaneously returns to its minimum and so on.

These transitions represent displacements of the dimer, described by $u$, which is defined by Eq. (9). Obviously, for a quantitative treatment we should know the values of the parameters. Some estimates have already been done [1] but this is still an open question. We should keep in mind that the model described in this paper is somewhat different from the original one, introduced in [1]. For example, Eqs. (11), (38) and (39) represent a relationship between some parameters and will help in their estimations. Anyway, the parameter selection requires further research and is not the topic of this work.

A careful reader may ask why only the function $\psi_1(\xi)$ has been studied in this paper. There are three solutions of (37) and, consequently, three possible functions $\psi(\xi)$. However, it does not seem that the functions $\psi_2(\xi)$ and $\psi_3(\xi)$, corresponding to $a_{02}$ and $a_{03}$, are physically relevant. For example, one can easily show that

$$\psi_3(-\infty) = \psi_L, \qquad \psi_3(+\infty) = \psi_{max}, \tag{53}$$

which could be expected. Obviously, the soliton $\psi_1(\xi)$, corresponding to the deeper energy minimum, is more relevant. Also, we can show that

$$\psi_2(-\infty) = \psi_R, \qquad \psi_2(+\infty) = \psi_L, \tag{54}$$

which may be physically relevant only for $\sigma < \sigma_0$. Therefore, we study in some more details only one out of the three possible solutions as this one is physically important. Also, we believe that the solution $\psi_1(\xi)$ is relevant for triggering mechanism, which will be explained in the last section.

We do not study the case when only one real solution of (37) exists. Such the case corresponds to the positive $b$, which does not have physical meaning, as was explained above.

Due to the large uncertainty in the values of the parameters $A$ and $B$ of the potential (4), we concentrate on the characteristics of the soliton that do not depend on these parameters. An important example is the wave velocity. After a simple algebra with Eqs. (11), (13), (36) and (42), it can be expressed as

$$v^2 = \frac{kl^2}{m + \frac{\gamma^2}{18a_0^2 A}}. \tag{55}$$

We consider the solution $a_0 = a_{01}$ given by Eqs. (44), (20) and (21).

According to the function $\psi_1(\xi)$ given by (49) we can write

$$\frac{3a_0}{\rho}\sqrt{1 - 3a_0^2}\,\xi = \frac{3a_0}{\rho}\sqrt{1 - 3a_0^2}\,\kappa(x - vt) \equiv \chi(x - vt). \tag{56}$$

Hence,

$$\chi = \frac{3a_0 \kappa A \sqrt{1 - 3a_0^2}}{\gamma \omega}. \tag{57}$$

Obviously, $\chi$ determines the solitonic width $\Lambda$ through [11]

$$\chi = \frac{2\pi}{\Lambda}. \tag{58}$$

It is convenient to write $\Lambda$ as an integer of dimer's length $l$, i.e.

$$\Lambda = Nl. \tag{59}$$

All this brings about the following expression for $v$:

$$v = \frac{3a_0 ANl\sqrt{1-3a_0^2}}{2\pi\gamma}. \tag{60}$$

We can eliminate $A$ from (55) and (60) and obtain

$$mv^2 + C_0 v - kl^2 = 0 \tag{61}$$

where

$$C_0 = \frac{Nl\gamma\sqrt{1-3a_0^2}}{12\pi a_0}. \tag{62}$$

Hence, the expression for $v$ reads

$$v = \frac{C_0}{2m}\left(-1 + \sqrt{1 + \frac{4mkl^2}{C_0^2}}\right). \tag{63}$$

It is obvious that the speed $v$ does not depend on $A$ and $B$ but does depend on some other parameters. Using (62) one can see that $v$ is the decreasing functon of viscosity $\gamma$. To estimate $v$ we can use [1]: $m = 11 \cdot 10^{-23}$ kg and $\gamma = 5.3 \cdot 10^{-11}$ kg/s but the values of $N$ and $k$ are not known. Also, of special importance is the intensity of electric field $E$ as this determines $a_0$ through $\sigma$. Namely, Eq. (12) can be written as

$$\sigma = \frac{qE}{A\sqrt{\frac{A}{B}}} = \frac{qdE}{dA\sqrt{\frac{A}{B}}} = \frac{pE}{dA\sqrt{\frac{A}{B}}}, \tag{64}$$

where $p \equiv qd$ is the longitudinal component of the electric dipole moment and the distance between the centres of positive and negative charges in any dimer is $d \approx 4$nm [5]. Therefore, the estimates of $E$, $A$ and $B$ are of crucial importance. An attempt to estimate these parameters is in progress even though some estimates were done in [1]. Some values of the velocity $v$ for $N = 30$ and four values of $k$ are given in Table 2 for different values of $\sigma$ (different strengths of the intrinsic electric field $E$).

Table 2

| $\sigma$ | 0 | 0.1 | 0.2 | 0.3 | $\sigma_0$ | |
|---|---|---|---|---|---|---|
| $v(\text{m/s})$ | 18.9 | 15.6 | 12.8 | 10.2 | 6.3 | $k = 0.1\,\text{N/m}$ |
| $v(\text{m/s})$ | 92.1 | 76.6 | 63.4 | 50.5 | 31.5 | $k = 0.5\,\text{N/m}$ |
| $v(\text{m/s})$ | 179.2 | 150.2 | 125.1 | 100.1 | 62.8 | $k = 1\,\text{N/m}$ |
| $v(\text{m/s})$ | 341.4 | 290.0 | 243.9 | 197.0 | 124.8 | $k = 2\,\text{N/m}$ |

## 4. Numerical solutions

In order to test the described analytical procedure and to study behaviour of the soliton for those parameters of the MT for which analytical solutions do not exist, we have solved Eq. (10) numerically. The standard shooting method with the 4$^{\text{th}}$ order Runge-Kutta integrator was used. Particular solutions are determined by their asymptotic behaviour and are centred at $x = 0$. The numerical step was $10^{-5}$.

Numerical solutions were first compared to the analytical solutions obtained by METHF method. Excellent agreement shown in Fig. 4 validates our analytical approach. We further generated numerical solutions of Eq. (10) for those parameters $\alpha$ for which our analytical procedure does not give solutions. To enable a direct comparison, we kept the parameters $\rho$ and $\sigma$ constant and again considered dimers within a MT with the intrinsic electric field zero ($\sigma = 0$) and $\rho = 1$, and with a non-zero electric field ($\sigma = 0.3$) and $\rho = 1$. The results are shown in Fig. 5 and Fig. 6, respectively. The values for $a_{01}$ were calculated according to (37).

As the parameter $\alpha$ multiplies only a derivative in Eq. (10) it does not influence asymptotic solutions. These are determined by $\sigma$ only. For small absolute values of $\alpha$ the soliton is narrow and has a shape similar to tanh function, meaning that the transition between the initial and the final position of the dimer is fast and smooth. On the other hand, for big $|\alpha|$ the dimmer experiences oscillation before it stabilizes in the final value. The period and amplitude of oscillations increase with $|\alpha|$. These results indicate that there is an optimal parameter $\alpha$ for a given process to occur.

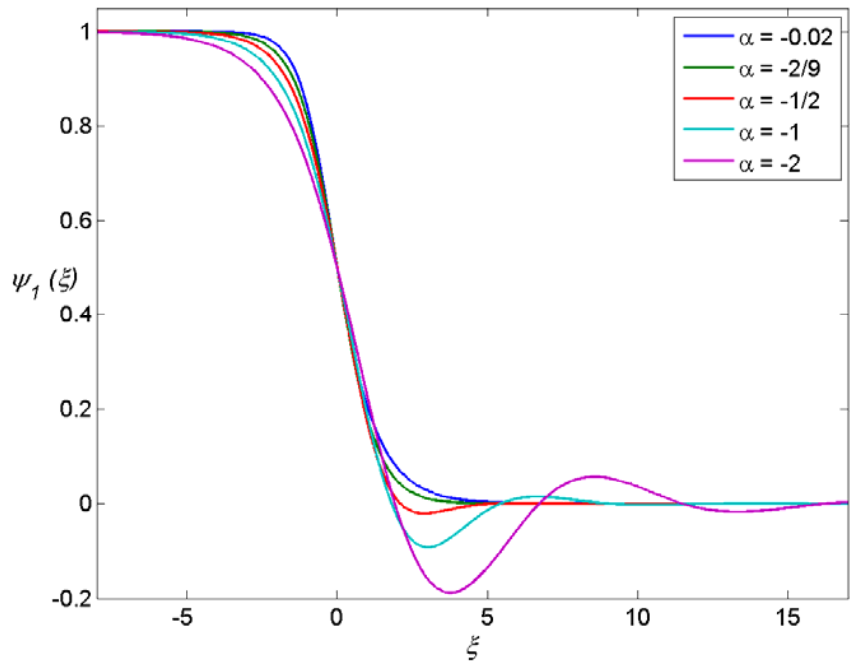

**Fig. 5.** The function $\psi_1(\xi)$ for $\rho = 1$ and $\sigma = 0$ for different values of $\alpha$

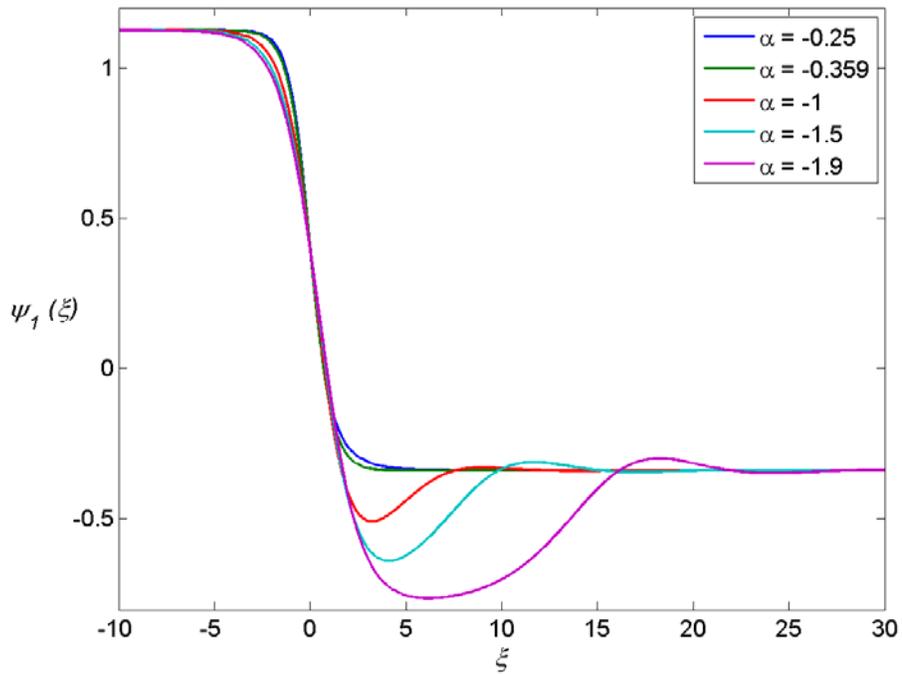

**Fig. 6.** The function $\psi_1(\xi)$ for $\rho = 1$ and $\sigma = 0.3$ for different values of $\alpha$

## 5. Conclusions

In this paper we thoroughly revisited an earlier work [1] dedicated to nonlinear dynamics of tubulin dimers within the same PF of a stable cellular MT. The basis of the model is the quartic nonlinear potential describing the mean field of chemical bonds between neighbouring tubulin dimers. This potential does dominantly depend on longitudinal displacements $u_n$ of the tubulin dimers along MT axis, as shown by Eq. (1).

Inclusion of dipole-dipole interaction between tubulin dimers as well as the impact of viscous damping of cytosol has led to dimensionless nonlinear ODE (10). The effective potential (15) was discussed respecting the parameter $\sigma$ which is proportional to the magnitude of intrinsic electric field created by MT itself.

To solve Eq. (10) we have relied on METHF method [6-10]. The physically tractable solutions have the shape of anti-kink soliton (49), well known in nonlinear ferroelectrics and ferromagnetics. These solutions for the two values of $\sigma$ are compared with the numerical solutions and a perfect agreement was demonstrated.

It might be important to point out that the Hamiltonian is only what was taken from Ref. [1]. In addition, the way how the viscosity term was introduced in the dynamical equation of motion is widely accepted and can be found in many references. The procedure explained in this paper is completely different from the one in Ref. [1]. It turned out that the METHF method is both elegant and useful. It enables us to estimate the maximum of electric field $E$. Namely, the highest value of $\sigma$ is $\sigma_0$, given by (21). According to Eq. (64), we see that this is proportional to $E$. As the values of $p$, $d$, $A$ and $B$ can be found in literature we can calculate the maximum of $E$. This is not the topic of this paper as we believe that the values of the parameters are still an open problem, requiring further research.

It was mentioned above that the crucial ODE for $\psi$ in this paper and in Ref. [1] are equal only if $\alpha = -1$, which certainly means that this approach is more general.

One can say that the final result of the both papers, this one and Ref. [1], is the function $\psi(\xi)$. We showed only the function $\psi_1(\xi)$ describing dynamics of MT when the system is in the deeper minimum of the potential well. This is shown in Figs. 4-6. There is one very important advantage of the method applied here over the one used in Ref. [1]. If we look at Eq. (3.11) of Ref. [1] we see that $\psi(\xi)$ does not depend neither on $\rho$ nor on $\sigma$, which is a big problem. In fact, Eq. (3.18) was an attempt to introduce the dependence of electric field but it turned out that the term with $\sigma$ was negligible. As for $\psi_1(\xi)$, existing in this paper, it depends on both parameters as $a_{01}$ depends on $\sigma$. This is the big advantage as $\rho$ and $\sigma$, being proportional to viscosity and internal electric field, are internal parameters and we certainly expect that the final result, i.e. dynamics of MT, should depend on them. In addition, the expression for $\psi_1(\xi)$ shows that the solitonic width is proportional to viscosity, which is physically plausible. Also, Figs. 4-6 show how $\sigma$ affects the final result.

Kink velocities are estimated in both in this paper and in Ref. [1]. We can see that these estimations are different. The estimated values of solitonic speed here are in the range from about 10m/s to a few hundreds of m/s. This interval may be comparable to the speed of nerve axon potential, spanning to $30 \text{m/s}$, but is much greater than the speed of MT motor protein which is, in a case of kinesin, around $1 \mu\text{m/s}$, or ionic waves within the cell (a few $\mu\text{m/s}$).

We expect that these stable solitons could be elicited for example by the energy released in hydrolysis of tubulin-nucleotide GTP, or by the interaction with other MT associated proteins.

We here offer the ideas of how these solitons could be utilized by a cell for some important mechanisms underlying its functional dynamics. First, it is plausible to expect that anti-kink soliton reaching the MT tip could cause the detachment of last tubulin dimer. This is one of triggering mechanisms responsible for depolymerisation of MTs.

Second, it is well known that MTs serve as a "road network" for motor proteins (kinesin and dynein) dragging different "cargos" such as vesicles and mitochondria to different sub-cellular locations. Kinesin motors move along MT towards MT plus-end and dynein motors move in opposite direction. Both classes of motors are present on the same "cargo" [12] and it is still unclear which mechanism decides which motor will take the action and drag the cargo in the proper direction. We argue that the cell's compartment which needs the specific cargo will launch the soliton of above kind sending it as a signal along the closest MT. The soliton will activate proper motors being close to MT along which the soliton propagates. For example, if the need for mitochondria is expressed in the regions close to the cell membrane, the solitons created in these MTs will propagate down MTs towards cell nucleus. These solitons will turn on kinesin motors in order to attach to the same MTs and to carry mitochondria towards MT plus-ends which are close to cell membrane. It might be useful to mention that some similar ideas were elaborated earlier [13,14].

**Appendix**

A polynomial

$$x^3 + px + q = 0, \quad p < 0 \tag{A.1}$$

has three real roots for $D < 0$ or one real and two complex conjugate roots for $D > 0$, where [15]

$$D = \frac{q^2}{4} + \frac{p^3}{27}. \tag{A.2}$$

For the negative $D$ the solutions of (A.1) are [15]

$$x_k = 2\sqrt[3]{r} \cos\left(\frac{\varphi + 2k\pi}{3}\right), \quad k = 0, 1, 2, \tag{A.3}$$

where

$$r = \sqrt{\frac{-p^3}{27}} \qquad (A.4)$$

and

$$\cos\varphi = -\frac{q}{2r}. \qquad (A.5)$$

For the positive $D$ the solution of (A.1) is [15]

$$x = -\frac{\sqrt{-\frac{p}{3}}}{\sin(2\varphi)} \qquad (A.6)$$

where

$$\tan\varphi = \sqrt[3]{\tan\left(\frac{w}{2}\right)} \qquad (A.7)$$

and

$$w = \arcsin\left[\frac{2}{q}\left(-\frac{p}{3}\right)^{\frac{3}{2}}\right]. \qquad (A.8)$$

**Acknowledgements**

This research was supported by funds from Serbian Ministry of Sciences, grants III45010 and OI171009.

L. Kavitha gratefully acknowledges the financial support from UGC in the form of major research project, BRNS, India in the form of Young Scientist Research Award and ICTP, Italy in the form of Junior Associateship.

S. Zdravković gratefully acknowledges the hospitality of the Abdus Salam International Centre for Theoretical Physics, Trieste, Italy, where a big part of this work was done.